\documentstyle[aps,prb,manuscript]{revtex}

\begin{document}
\draft

\title{Transition Temperature of Josephson Junction Arrays with Long-Range
 Interaction}
\author{H.R. Shea and M. Tinkham}
\address{Department of Physics and Division of Engineering and Applied Sciences,
Harvard University, Cambridge, MA 02138}
\date{\today}
\maketitle

\begin{abstract}
We report measurements of the dependence on magnetic field and array 
size of the resistive transition of Josephson junction arrays with 
long-range interaction.  Because every wire in these arrays has a 
large number of nearest-neighbors (9 or 18 in our case), a mean-field 
theory should provide an excellent description of this system.  Our 
data agree well with this mean-field calculation, which predicts that 
$T_c$  (the temperature below which the array exhibits macroscopic 
phase coherence) shows very strong commensurability effects and scales 
with array size.

\end{abstract}
\pacs{74.50.+r}

 We report an experimental investigation of ordered Josephson junction 
 Arrays with Long-Range Interaction (ALRI), of the sort originally 
 proposed in the disordered limit by Vinokur \emph{et al.} \cite{vin} 
 Although such arrays had been fabricated by Sohn \emph{et al.} 
 \cite{sohn2}, the samples used in the present work for the first time 
 have low enough critical currents and hence low enough screening to 
 be in the regime well described by existing theoretical models 
 \cite{sohn1,chandra}.
 
 These arrays consist of two perpendicular sets of $N$ parallel 
 superconducting wires, coupled by Josephson junctions at every point 
 of crossing (see Fig.\ \ref{fig1}).  In this geometry, any 
 horizontal (vertical) wire is nearest neighbor to \emph{all} vertical 
 (horizontal) wires, and next-nearest neighbor to \emph{all} other 
 horizontal (vertical) wires.  Hence we term the interaction 
 long-range.  The number of nearest neighbors in these arrays is equal 
 to the array size $N$.  This is in sharp contrast to standard 
 (short-range interaction) $2D$ arrays where the number of nearest 
 neighbors (typically 4 or 6) is independent of array size.
 
 Arrays with long-range interaction were first proposed as a model for 
 a spin glass for the case where the wires are positionally disordered 
 and a sufficiently strong perpendicular magnetic field is applied 
 \cite{vin}.  More recently, Chandra \emph{et al.} \cite{chandra} have 
 shown that even for an \emph{ordered} array, glassy behavior is 
 expected in a very weak field (less than one flux quantum per row).  
 The equivalent of ``spins'' in these ALRI are the phases of the 
 superconducting wires.  Above a transition temperature $T_{c}$, the 
 phases of the wires are uncorrelated.  However when the array is 
 cooled below $T_{c}$, a transition to a macroscopically phase coherent 
 state is predicted to occur.
 
For an ordered array with long-range interaction in the limit of 
negligible screening, Sohn \emph{et al.}\cite{sohn1} have performed a 
mean-field analysis and computed the transition temperature 
$T_{c}^{MF}(f)$ as a function of the applied field and array size.  
Because each wire has a large number of nearest-neighbors, a 
mean-field theory using the phase of each wire as a classical 
thermodynamic variable should provide a good description of this 
system.  At zero field they find $T_c={{N E_J(T=T_{c})}/{2k_B}}$, where 
$E_{J}(T)=\hbar i_{c}(T)/2e$ and $i_{c}(T)$ is the (unfluctuated) 
critical current of a single junction at temperature T. Note the 
unusual result that $T_{c}$ should scale with the \emph{size} of the 
array.  To keep $T_{c}$ of the array well below $T_{c}^{wire}$, one 
requires $ N \hbar i_c^0 \ll 4ek_BT_c^{wire}$, where $i_{c}^{0}$ is 
$i_{c}(T=0)$.  The computations of Refs.\ [1], [3], and [4] only hold 
in the limit of negligible screening, where the array as a whole 
screens much less than one flux quantum $\Phi _0$, and in the limit 
when phase gradients along the wires due to current flow are much less 
than phase drops at the junctions.  The former condition can be 
written as $ N^2 L_g i_c^0 \ll \Phi _0 $, where $L_{g}$ is the 
geometric inductance of a cell in the array.  The latter condition can 
be expressed as $i_{c}^{junction} \ll i_{c}^{wire}$.  These three 
inequalities place very strong limits on the magnitude of $i^{0}_{c}$ 
for given $N$.
	
In the experimental work of Sohn and co-workers, ${{N\hbar i_c^0} / 
{4ek_BT_c^{wire}} }\approx 300$ and $N^{2}L_{g}i^{0}_{c}/\Phi _0 
\approx 10^{3}$.  Hence their samples were not in the regime 
defined by the above mentioned theories.  We present here the first 
measurements of ALRI with critical currents small enough 
(of order 5 nA) to be in the limit of extremely weak screening, and to 
have an array $T_{c}$ well below the wire critical temperature.  Our 
data show impressive agreement with the mean-field theory, including 
extremely strong commensurability effects.

The samples consist of 0.25 $\mu$m wide Al wires 
($T_{c}^{wire}\approx 1.7$ K) connected by Al-AlO$_{x}$-Al junctions, 
fabricated as follows.  A grid-like pattern of lattice constant 2 
$\mu$m is defined using electron-beam lithography on a PMMA-coated Si 
wafer.  A three-angle shadow evaporation technique is used to deposit 
both sets of wires sequentially without breaking vacuum, using only 
the single lithography step.  The evaporations are done at 
$45^{\circ} $ to the substrate surface, but at different orientations 
with respect to the patterned channels.  30 nm of 99.999\%
pure Al are evaporated in the direction of one set of wires (the 
``horizontal'' set).  Al accumulates on the substrate only along those 
horizontal wires because the PMMA shadows the ``vertical'' wires.  150 
mTorr of O$_{2}$ is bled into the chamber, and an oxygen plasma is 
ignited for 20 minutes to grow an AlO$_{x}$ barrier.  After pumping 
out the O$_{2}$, the sample is rotated so that the second and third 
evaporations (30 nm of Al each) are done in the direction of the 
``vertical'' wires, going ``up'' for the second evaporation and 
``down'' for the third, to ensure that the vertical lines are 
continuous where they ``climb'' over the horizontal wires.  A liftoff 
completes the process.  This shadow evaporation technique yields very 
high quality underdamped junctions which are a major improvement over 
those from the previous fabrication technique.\cite{sohn2}

The typical single junction resistance is $R_{N}^{JJ}=$70 k$\Omega$ 
which corresponds\cite{baratoff} to an unfluctuated critical current 
$i_{c}^{0}$ of 5.6 nA, or $E_{J}(T=0)/k_{B}=0.13$ K. Junction uniformity, 
measured from single junctions co-fabricated with the arrays, is 
approximately $\pm 15$\%.  The lead configuration is shown in the 
inset of Fig.\ \ref{fig2}.
Current is injected in the first wire of one set, and extracted from 
the last wire of that same set.  We report data on two arrays: one 
consisting of $9\times 9$ wires ($8\times 8$ cells) and the other of 
$18\times 18$ wires ($17\times 17$ cells).

The arrays are cooled to 315 mK in a $^{3}$He cryostat within a 
double $\mu$-metal shield which reduces the stray field to less than 
50 mG. (A field of 5.2 G corresponds to $f \equiv \Phi _{cell}/\Phi 
_{0}=1$ for our 2 $\mu$m spacing).  Temperature stability is better 
than 3 mK below 2 K. A small magnetic field is applied perpendicular 
to the array using a solenoid surrounding the vacuum can of the 
cryostat.  Screening by the array can be neglected because 
$i_{c}^{0}$ is so small.  Quantitatively, the ratio of the maximum 
flux screened by the array to the flux quantum is much less than one: 
$N^{2}L_{g}i^{0}_{c} /\Phi _{0} \approx 3\times 10^{-3} \ll 1$, for 
$N=18$ and where $L_{g}\approx 4$ pH is the geometric inductance of a 
single cell in the array, modeled as a superconducting square 
washer.\cite{jaycox} Considerable care was taken to ensure that the 
arrays are well shielded from RF and microwave radiation by the use 
of a shielded room, room-temperature low-pass LC $\Pi$-filters, cold 
resistors, and cold microwave filters.\cite{martinis}

The current-voltage (I-V) curves for single junctions co-fabricated 
with the arrays do not show a well-defined critical current at 0.3 K 
because $E_{J}<k_{B}T$, and hence a finite resistance is observed for 
all bias currents.  The arrays on the other hand, consisting of many 
junctions in parallel, do show, at least at the lowest temperatures, a 
well-defined critical current and strong hysteresis, as expected from 
underdamped SIS junctions.  Fig.\ \ref{fig2} shows an I-V curve for 
the $17\times 17$ array at $f=0$.  The two jumps in voltage to 
$2\Delta /e$ and $4\Delta /e$ (where $\Delta$ is the superconducting 
gap) correspond to all the junctions connected to one, then the other, 
of the current injection wires going normal.  The unfluctuated 
zero-temperature array critical currents $I^{0}_{c}=Ni^{0}_{c}$ 
($\sim$60 nA for the $9\times 9$ wire array, $\sim$100 nA for the 
$18\times 18$ wire array) are so small that the measured $I_{c}$ will 
be significantly less than $I_c^0$ due to thermal fluctuations.  The 
measured $I_{c}$ actually corresponds to a jump from a finite-voltage 
(of order 1 $\mu$V) phase-diffusion branch \cite{phasediff} to 
$2\Delta /e$ at a current value which is affected by damping as well 
as $E_{J}$ and $T$.

We therefore focus instead on the differential resistance 
$R_{d}=dV/dI$ (at zero dc bias) as a function of field and temperature 
since it should be a better measure of the zero-current phase coupling 
of the array (and hence $T_{c}$).  $R_{d}$ is measured using a PAR 124 
lock-in amplifier at 15.6 Hz with an excitation current of 0.2 nA 
(corresponding to $\sim I_{c}/10$).  Fig.\ \ref{fig3} shows $R_{d}$ 
vs.  $f \equiv \Phi _{cell}/\Phi _{0}$ plots for several temperatures 
from 0.3 K to 1.6 K for the $17\times 17$ cell sample.  The curves are 
\emph{not} offset.  Because $R_{d}(f)$ is periodic in $f$ and 
symmetric around $f=1/2$, we only plot $R_{d}(f)$ for $f$ ranging from 
0 to 1/2.  $R_{d}(f)$ displays minima at \emph{all} commensurate 
fields where $f=p/q$, $p$ and $q$ being integers smaller than $N$.  
The $f=1/17$ and $f=1/16$ states are not clearly resolved but all 
other commensurate states are clearly present (such as, for instance, 
all other multiples of 1/17, like 2/17 and 3/17).  \emph{All} the 
measured positions of the resistance minima are within less than 
$10^{-4}\times \Phi _{0}$ from the ideal computed commensurate field 
values.  We observe very similar behavior for the $8\times 8$ cell 
array, with resistance minima at $f=1/8, 1/7,\ldots$ It is a 
characteristic feature of ALRI that such strong and detailed structure 
in the $R_{d}(f)$ curve is visible.  Standard $2D$ arrays and wire 
networks do not exhibit such richness of structure because they do not 
have the long-range order needed to support a stable vortex 
superlattice with such a large lattice constant (e.g.\ 17 cells).

The deepest resistance minima occur at the most strongly commensurate 
states: $f=0$, 1/2, 1/3, 1/4.  The shape of the $R_{d}(f)$ curve is 
very similar near all of these states (see Fig.  \ref{fig3}).  The 
full widths of the resistance dips (i.e.\ the field intervals between 
local maxima on either side of the dips) scale as $1/q$, with 
$q=1,2,3\ldots$ Near integer $f$ (e.g.\ $f=0,1,2\ldots$) where $q=1$, 
the resistance increases smoothly from $f=n$ until $f=n \pm 
[1/(N-1)]$, i.e.\ the first adjacent commensurate state, giving a 
modulation-free half-width of $1/(N-1)$. Corresponding behavior 
occurs near other strongly commensurate states.

$R_{d}(f)$ was measured for 20 temperatures between 0.315 K and 1.8 K, 
of which 11 are shown in Fig.  \ref{fig3}.  As the temperature is 
increased, $R_{d}(f)$ increases and the relative amplitudes of the 
resistance oscillations decrease until at higher temperatures (but 
with the wires still superconducting) $R_{d}(f)$ saturates at the 
normal state resistance of the array $R_{N}^{array} = 2 R_{N}^{JJ}/18$. 
In order to extract $T_{c}(f)$ from the $R_{d}(f)$ vs.  $T$ data, we 
make use of the finite width of the resistive transition.  We 
\emph{define} the experimental $T_{c}$ using a resistive criterion;  
for each field value, $R_{d}$ is plotted vs.  $T$, and $T_{c}$ is 
taken to be the temperature at which $R_{d}$ interpolates to $\epsilon 
R_{N} $, and $\epsilon$ is a number between 0 and 1.  Automating this 
process produces the top two curves of Fig.\ \ref{fig4} of $T_{c}(f)$ 
for $\epsilon=0.5$ (top curve) and $\epsilon=0.375$ (middle curve).  
For values of $\epsilon$ between approximately 0.4 and 0.8, the 
inferred $T_c$ scales almost linearly with $\epsilon$.
 
The bottom curve is the result of a mean-field calculation of 
$T_{c}^{MF}(f)$ for a $17\times 17$ cell array.  $T_{c}^{MF}(f)$ is 
the temperature above which the order parameter $\eta _{i} \equiv 
\langle e^{i\phi _{i}} \rangle$ is equal to 0, where $\phi _{i}$ is 
the phase of the $i^{th}$ wire and the brackets denote a thermal 
average.  There are no free parameters in this calculation, which 
consists of using an efficient scheme to find the largest eigenvalue 
of a $17\times 17$ matrix given by Eq.\ (19) of Ref.\  [3] for each of 
one thousand field values shown.  The eigenvalue problem is solved 
assuming a temperature-independent $E_J$, and $T_c^{MF}$ is finally 
corrected to account for $E_J(T)$, which varies by $\sim$ 30 \% over
the temperature range of interest.

The data and mean-field theory curves are in good agreement, both for 
the $17\times 17$ cell array in Fig.\ \ref{fig4}, and for the $8\times 
8$ cell array (not shown).  The maxima in the experimental $T_{c}(f)$ 
obviously occur at commensurate fields to the same high accuracy as 
the minima in the resistive data do, since the critical temperature 
was extracted from the $R_{d}(f)$ curves.  The mean-field theory also 
predicts local maxima in $T^{MF}_{c}$ at all commensurate fields: we 
find that the positions of the clearly discernible maxima in the 
experimental $T_{c}(f)$ and $T^{MF}_{c}$ agree to better than one part 
in $10^{4}$.  As can be seen in Fig.\ \ref{fig4}, the lower resistance 
criterion gives better quantitative agreement with the mean-field 
theory $T_c^{MF}$, which is always below the experimental $T_c$ (and 
is defined slightly differently).  We cannot use a resistive criterion 
of less than $\epsilon \approx 0.37$ over the whole field range 
because at low temperatures, for $f\neq0$, the array resistance 
saturates at a non-zero value (up to $\approx$ 3 k$\Omega$ for 
incommensurate $f$), probably due to macroscopic quantum tunneling of 
the phases.
 
In order to compare the experimental $T_{c}$'s of the $18\times 18$ 
wire and $9\times 9$ wire arrays, we must first account for the 
temperature dependence of $E_{J}$ in order to obtain $T_{c}^{*}$, the 
transition temperature one would observe if $E_{J}$ were constant and 
the same for both arrays.  For $f=0$ we then obtain $T_{c,18}^{*} / 
T_{c,9}^{*} =1.9$, using $\epsilon =0.5$ to determine $T_{c}$ for both 
arrays. This is very close to the theoretical value of $18/9=2$, indicating 
that $T_{c}$ does indeed scale with system size $N$.

In the absence of screening we can write the following simple 
expressions for the phases of each wire at $x=ja$ and $y=ka$ at zero 
applied current:
$$\varphi _k^H=\varphi _{0}^H+2\pi fNk$$
$$\varphi _j^V=\varphi _{0}^V+2\pi fkj .$$
$\varphi ^{H}_{k}$ is the phase of the $k^{th}$ horizontal wire 
(constant along the wire) and $\varphi ^{V}_{j}$ is the phase of the 
$j^{th}$ vertical wire (depends linearly on the position y along the 
wire).  $a$ is the lattice constant and the vector potential has been 
chosen as ${\bf A}=fx\Phi _0/ a^2{\bf\hat y}$ .
 The only free parameter is $\Delta \varphi 
^{HV}_{0}=\varphi ^{H}_{0} - \varphi ^{V}_{0}$.  The 
system energy $E$ is
 $$E=-\sum\limits_{k,j=0}^{N-1}
{\cos \left( {\varphi _k^H-\varphi _j^V} \right)}$$
The ground state energy is found by minimizing $E$ numerically as a 
function of $\Delta \varphi ^{HV}_{0}$ for each field.  Once $\Delta 
\varphi ^{HV}_{0}$ is found, all the phase differences are then 
determined.  The local extrema of both $-E_{min}(f)$ and 
$T^{MF}_{c}(f)$ occur at exactly the same fields, with very similar 
relative amplitudes, indicating that the above simple expressions for 
the phases of the wires do indeed describe the phases very accurately.

It is very difficult to probe the glassiness of this system using 
transport measurements because the phases unlock as soon as a small 
transport current is applied.  Even though the arrays are biased well 
below $I_c$, a finite voltage develops across the system because of 
 phase diffusion. Phase diffusion is unavoidable 
in the small (i.e.\ low-capacitance) and weak junctions required to 
conform to the model conditions.  Since the phases are 
evolving as $\langle {d\varphi /dt}\rangle =2eV_{dc}/\hbar $, they 
cannot lock.  Hence individual metastable states, the presence of 
which would confirm the presence of a glass, cannot be probed using 
our transport technique.  For instance, we observe the same $I_c$ at 
every field cool, while trapping into different metastable states 
should give a range of measured critical currents.  Similarly, the 
$T_c$ we measure reflects an average over many configurations and thus
reveals very little about the glassiness of the array.

In conclusion, we have fabricated Josephson junction arrays with 
long-range interaction and extremely weak critical currents.  A 
mean-field theory provides an excellent description of this system 
because every wire has a large number of nearest-neighbors (9 or 18 
for the arrays presented here).  Our data for $T_{c}(f,N)$ are in very 
good agreement with the mean-field calculation: we find that 
$T_{c}(f=0)$ scales with system size and observe very strong 
commensurability effects.  The array differential resistance at zero 
dc bias exhibits minima at \emph{all} commensurate fields, displaying 
far more complex, but well understood, structure than standard 2D 
arrays or wire networks.

We wish to thank R.J.\ Fitzgerald and M.A.\ Itzler for their extensive 
assistance with the measurements and analysis, and J.M.\ Hergenrother, 
D.\ Davidovic, and M.S.\ Rzchowski for their insightful comments.  
H.R.S.\ acknowledges support of NSERC of Canada and FCAR du Qu\'{e}bec 
fellowships.  This work was supported in part by ONR Grant No.\ 
N00014-96-1-0108, JSEP Grant No.\ N00014-89-J-1023, and NSF Grant No.\ 
DMR-92-07956.

\begin{figure}
 \caption{ Schematic drawing of a 2 wire by 3 wire array with 
 long-range interaction.  There are Josephson junctions at every 
 crossing point of the superconducting wires. }
\label{fig1}
\end{figure}

\begin{figure}
 \caption{Voltage-Current plot at 0.315 K in zero field of the 
 $17\times 17$ cell array.  The dashed line corresponds to sweeping 
 current up, the solid line to sweeping current down. There is a 
 finite slope at zero bias, too small to be seen on the graph.  The 
 inset is a schematic diagram of the lead configuration used for current 
 injection and voltage measurement.}
    \label{fig2}
\end{figure}

\begin{figure}
   \caption{Plot of zero-bias differential resistance of the $17\times 
   17$ cell array vs.\ normalized flux $f$, measured with a 0.2 nA ac 
   excitation, for selected temperatures.  The curves are \emph{not} 
   offset.  From the lowest to the highest curve, the temperatures 
   are: 0.417 K, 0.702 K, 0.797 K, 0.959 K, 1.047 K, 1.13 K, 1.212 K, 
   1.288 K, 1.39 K, 1.51 K, and 1.69 K. The local minima in resistance 
   occur within $10^{-4}\times \Phi _{0}$ of all the commensurate flux 
   values, i.e.  at \emph{all} $f=p/q$ where $p$ and $q$ are integers 
   between 1 and 17.}
\label{fig3}
\end{figure}

\begin{figure}
 \caption{Plot of the temperature $T_{c}$ corresponding to the onset 
 of macroscopic phase coherence vs.  normalized flux $f$ for the 
 $17\times 17$ cell array.  The top two curves (data) are computed 
 from the differential resistance vs.\ field data by using a resistive 
 transition criterion for $T_{c}$ of $0.5 R_N^{array}$ (top curve) and 
 $0.375 R_N^{array}$ (middle curve).  The lower curve is the result of 
 a mean-field calculation of $T_{c}^{MF}(f)$.}
\label{fig4}
\end{figure}

\end{document}